\documentclass[aps,prl,twocolumn,superscriptaddress,showpacs]{revtex4-1}
\usepackage{graphicx}
\usepackage{dsfont}
\usepackage[latin1]{inputenc}
\usepackage{version} 
\usepackage{color}
\usepackage{amsmath}
\usepackage{braket}
\usepackage{bm}
\usepackage{upgreek}
\usepackage{colortbl}
\usepackage{graphicx}
\usepackage{dcolumn}
\usepackage{bm}
\usepackage{bigints}
\usepackage{xcolor}  
\usepackage[pagebackref=false]{hyperref} 

\definecolor{jrp}{rgb}{1,0,0}

\definecolor{mjg}{rgb}{.08,.05,.8}

\definecolor{yyl}{rgb}{.8,.05,.08}

\newcommand{\delete}[1]{{}}

\hypersetup{
    bookmarks=true,         
    unicode=false,          
    pdftoolbar=true,        
    pdfmenubar=true,        
    pdffitwindow=false,     
    pdfstartview={FitH},    
    pdftitle={Electrically Protected Valley-Orbit States in Silicon},    
    pdfauthor={X. Mi},     
    pdfkeywords={keyword1} {key2} {key3}, 
    pdfnewwindow=true,      
    colorlinks=true,       
    linkcolor=red,          
    citecolor=blue,        
    filecolor=magenta,      
    urlcolor=cyan,           
}

\makeindex

\begin{document}

\title{Electrically Protected Valley-Orbit Qubits in Silicon} 
\author{X.~Mi}
\affiliation{Department of Physics, Princeton University, Princeton, New Jersey 08544, USA}
\author{S.~Kohler}
\affiliation{Instituto de Ciencia de Materiales de Madrid, CSIC, E-28049 Madrid, Spain}
\author{J.~R.~Petta}
\email{petta@princeton.edu}
\affiliation{Department of Physics, Princeton University, Princeton, New Jersey 08544, USA}
\date{\today}

\begin{abstract}
Electrons confined in Si quantum dots possess orbital, spin, and valley degrees of freedom (d.o.f.). We perform Landau-Zener-St\"uckelberg-Majorana (LZSM) interferometry on a Si double quantum dot that is strongly coupled to a microwave cavity to probe the valley d.o.f. The resulting LZSM interference pattern is asymmetric as a function of level detuning and persists for drive periods that are much longer than typical charge decoherence times. By correlating the LZSM interference pattern with charge noise measurements, we show that valley-orbit hybridization provides some protection from the deleterious effects of charge noise. Our work opens the possibility of harnessing the valley d.o.f. to engineer charge-noise-insensitive qubits in Si.
\end{abstract}

\maketitle

Environmental fluctuations are pervasive in biological and condensed matter systems \cite{Lestas_Nature_2010,Weissman_RMP_1988}. Electrical fluctuations, commonly referred to as charge noise, typically fall off with frequency $f$ with a spectrum that is close to 1/$f$, and can induce uncontrolled evolution of quantum systems \cite{Paladino_RMP_2014}. These fluctuations are often the leading source of decoherence for solid-state qubits. For example, early superconducting qubits were highly sensitive to charge noise \cite{Nakamura_Nature_1999,Ithier-PRB_2005}. Even semiconductor spin qubits are sensitive to electrical noise, which limits the fidelity of single spin rotations in isotopically enriched $^{28}$Si \cite{Yoneda_threenines_2017} and two-spin gate operations based on the exchange interaction \cite{Petta_Science,Dial_Charge_2013}. It is therefore of critical importance to reduce charge noise or find ways to mitigate its impact on qubit coherence. 

Developing electrically protected qubits has been a recurring theme in solid-state quantum computation, both for superconducting and semiconductor qubits. Superconducting qubits have improved performance when operated at ``sweet spots" where the qubit transition energy is first-order insensitive to the level detuning \cite{Vion_Science_2002}. Optimization of the ratio of the Josephson energy $E_{\rm J}$ to the charging energy $E_{\rm C}$ has led to the development of transmon qubits that are highly insensitive to charge noise \cite{Schrier_PRB_2008}.   To date, approaches to mitigating the detrimental impact of charge noise on semiconductor qubits include dynamic decoupling \cite{Dial_Charge_2013}, device operation at sweet spots \cite{Petersson_PRL_2010,Reed_PRL_2016,Martins_PRL_2016}, and hybrid qubits \cite{Kim_Hybrid_2014}, where higher lying states in the qubit energy level spectrum lead to flat bands with an energy separation that is largely insensitive to charge fluctuations.

Here we demonstrate a DQD charge qubit that utilizes valley-orbit mixing in Si to achieve a qubit transition energy with reduced sensitivity to charge noise over a wide range of parameter space. The bulk conduction band of Si has six equivalent valleys. Strain in Si/SiGe heterostructures partially lifts the six-fold valley degeneracy by raising the energy of the four in-plane valleys \cite{Schaffler_SiGe_Review}. The electric field at the quantum well interface hybridizes the two low-lying $\pm$$z$-valleys, yielding a valley splitting in the range of 10--300 $\mu$eV \cite{Borselli_ValleyAPL_2011,Mi_ValleyPRL_2017,Schoenfield_NatComm_2017}. In Si double quantum dots (DQDs), hybridization of orbital and valley degrees of freedom leads to valley-orbit couplings that exceed 10 $\mu$eV \cite{Schoenfield_NatComm_2017,Mi_ValleyPRL_2017}. We show here that valley-orbit coupling gives rise to hybrid valley-orbit states that have a transition frequency that only weakly depends on the DQD level detuning -- an attribute that protects the valley-orbit states from charge decoherence in a manner analogous to superconducting transmon qubits \cite{Koch_PRA_2007}.

\begin{figure}[t]
	\centering
	\includegraphics[width=1\columnwidth]{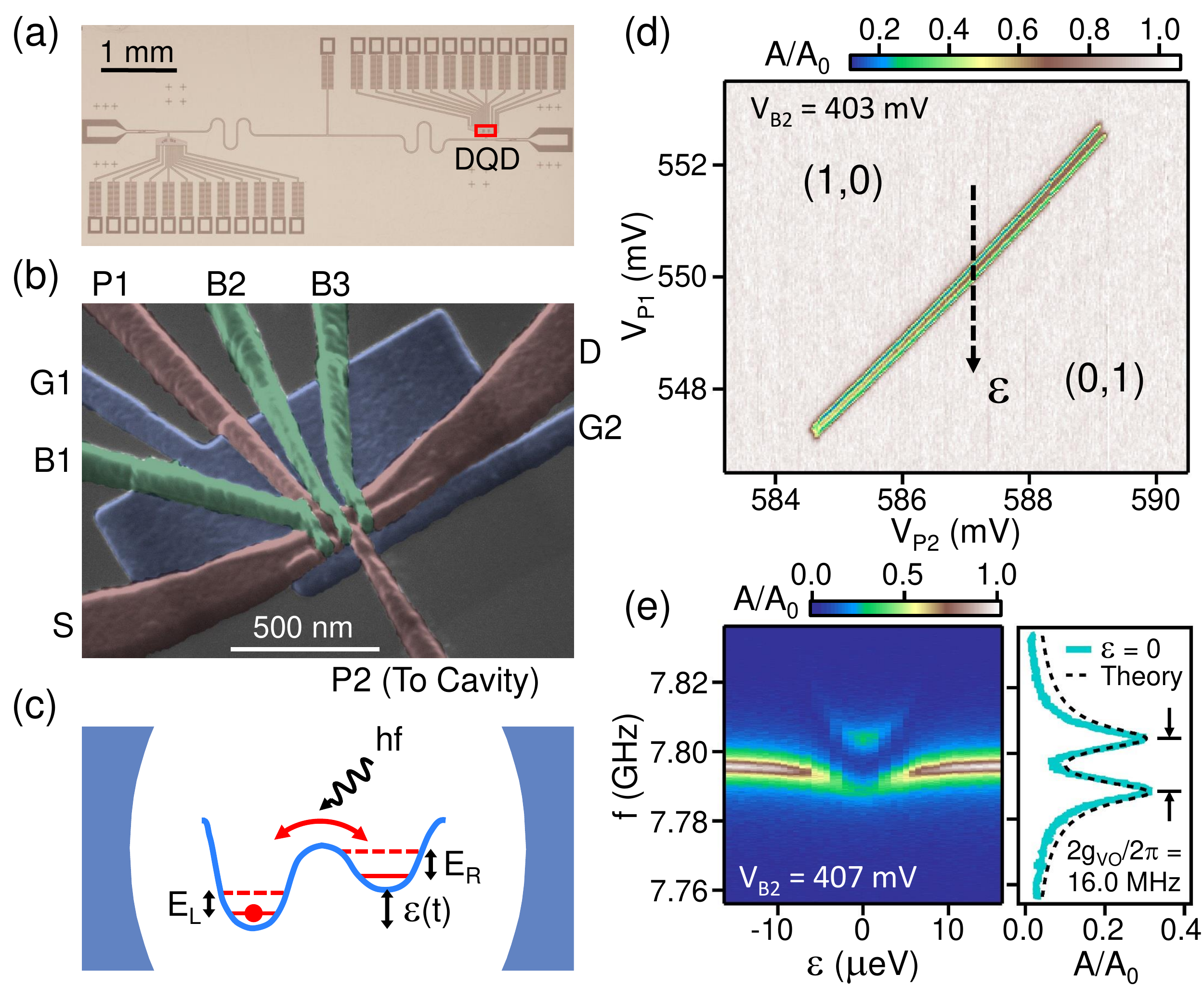}
	\caption{(a) Optical image of the superconducting cavity. (b) Tilted-angle false-color scanning electron microscope image of the DQD. (c) Experiment schematic: The DQD is probed by a cavity photon with energy $hf$ and interferometry is performed by periodically driving the detuning parameter $\epsilon (t)$. (d) Cavity transmission amplitude $A/A_0$ as a function of $V_\text{P1}$ and $V_\text{P2}$ near the $(1,0) \leftrightarrow (0,1)$ interdot transition. (e) Left panel: $A/A_0$ as a function of $\epsilon$ and $f$. Right panel: $A/A_0$ as a function of $f$ at $\epsilon = 0$.}
	\label{fig:1}
\end{figure}

\begin{figure*}[th]
\centering
\includegraphics[width=2\columnwidth]{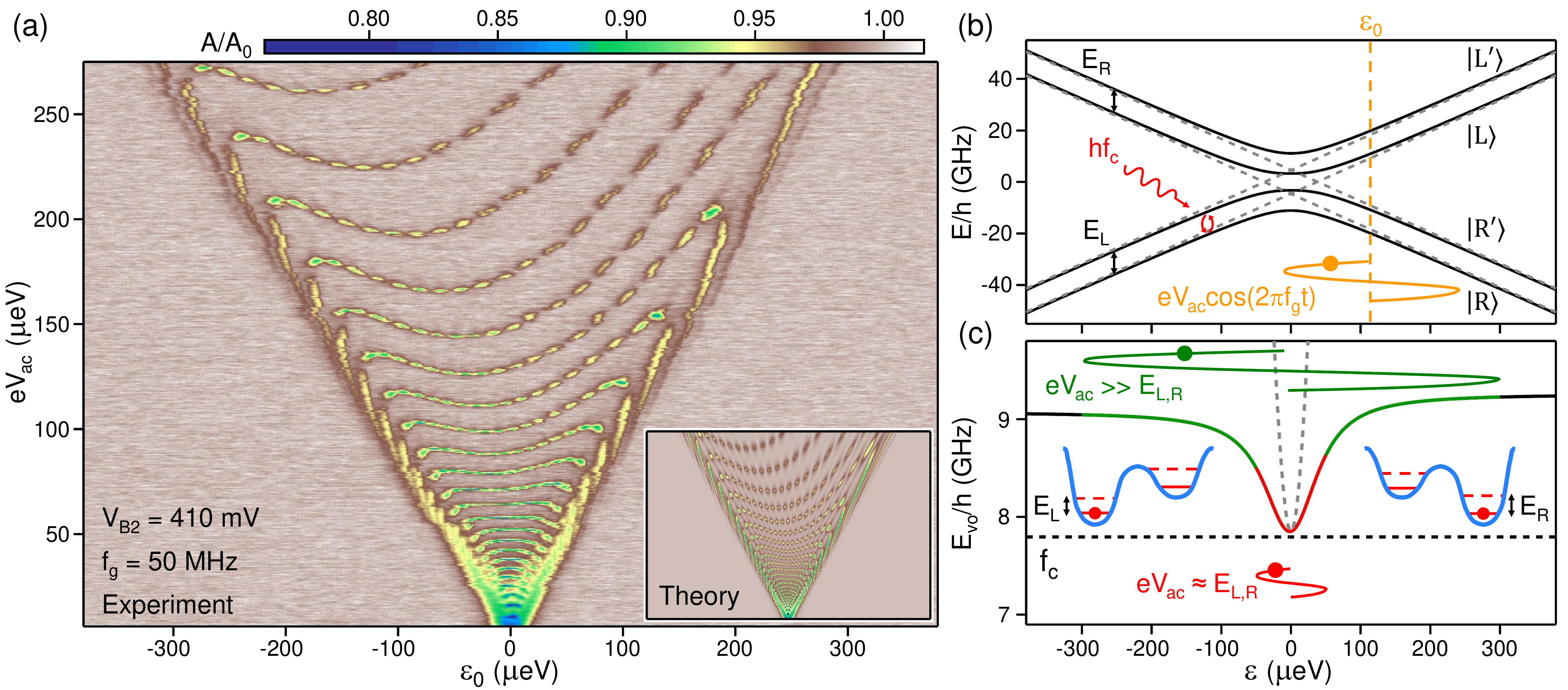}
\caption{(a) LZSM interference pattern obtained by measuring $A/A_0$ as a function of $\epsilon_0$ and $eV_\text{ac}$. Inset: Theoretical predictions plotted using the same color-scale and axes range as the experimental data. (b) DQD energy level diagram calculated with $E_\text{L} = 37.5$ $\mu$eV, $E_\text{R} = 38.3$ $\mu$eV, $t = 25.4$ $\mu$eV and $t' = 11.8$ $\mu$eV. The gray dashed lines show the energy levels with $t = t' = 0$. (c) Transition frequency between the two lowest energy states, $E_\text{VO} / h$, as a function of $\epsilon$. The cavity frequency $f_\text{c}$ is denoted with a black dashed line. The qubit-cavity detuning $E_\text{VO} / h - f_\text{c} \gg g_\text{VO} / 2 \pi$ and has a minimum of 55 MHz at $\epsilon = 0$. The red curve represents the detuning trajectory when $eV_\text{ac} \approx E_\text{L,R}$, which is confined to the parabolic parts of $E_\text{VO} (\epsilon) / h$ (shaded in red). The green curve represents the detuning trajectory when $eV_\text{ac} \gg E_\text{L,R}$, where the detuning trajectory samples flat parts of $E_\text{VO} (\epsilon) / h$ as well (shaded in green). For comparison,  the gray dashed line shows the transition frequency of a two-level charge qubit $E_\text{CQ}/h = \sqrt{\epsilon^2 + 4 t_\text{c}^2}/h$ with $t_\text{c} =$ 16.2 $\mu$eV. }
\label{fig:2}
\end{figure*}

The device used in this experiment consists of a Si/SiGe DQD that is electric-dipole-coupled to a microwave cavity having a center frequency $f_\text{c} = 7.796$ GHz and photon decay rate $\kappa/2 \pi = 3.3$ MHz [Figs.~\ref{fig:1}(a)--\ref{fig:1}(b)] \cite{Mi_Science_2016,Mi_ValleyPRL_2017}. The Hamiltonian governing the DQD charge states and valley d.o.f. is \cite{PhysRevB.94.195305}:
\begin{equation}
H_0 = 
\begin{pmatrix}
\epsilon/2+E_\text{L} & 0 & t & t' \\
0 & \epsilon/2 & -t' & t \\
t & -t' & -\epsilon/2+E_\text{R} & 0 \\
t' & t & 0 & -\epsilon/2
\end{pmatrix}.
\end{equation}
The Hamiltonian is written in a basis spanned by the left dot ground (excited) state $\ket{L}$ ($\ket{L'}$) and the right dot ground (excited) state $\ket{R}$ ($\ket{R'}$). Here $E_\text{L}$ ($E_\text{R}$) is the valley splitting of the left (right) dot, $t$ ($t'$) is the intra-valley (inter-valley) tunnel coupling and $\epsilon$ is the DQD level detuning [Fig.~\ref{fig:1}(c)]. Intra-valley tunnel coupling is equivalent to the interdot tunnel coupling $t_c$ in single valley materials \cite{Petersson_PRL_2010}.

Values of $E_\text{L}$, $E_\text{R}$, $t$ and $t'$ are measured through the dispersive interaction between the DQD and cavity photons \cite{Mi_ValleyPRL_2017}. We first identify the $(1,0) \leftrightarrow (0,1)$ interdot charge transition by measuring the normalized cavity transmission amplitude $A/A_0$ as a function of the left (P1) and right (P2) plunger gate voltages $V_{\rm P1}$ and $V_{\rm P2}$ [Fig.~\ref{fig:1}(d)] \cite{Mi_Science_2016}. Here $(N_1,N_2)$ denotes a charge state with $N_1$ ($N_2$) electrons in the left (right) dot. Vacuum Rabi splitting with a frequency $2g_\text{VO} / 2 \pi = 16.0$ MHz [Fig.~\ref{fig:1}(e), left panel] is observed in the cavity transmission spectrum at the $\epsilon = 0$ sweet spot \cite{Mi_Science_2016,Strong_Coupling_CooperPair,Petersson_PRL_2010}. A charge decoherence rate $\gamma_0 / 2 \pi = 4.1$ MHz is extracted using microwave spectroscopy \cite{Supplement}, indicating the device is in the strong-coupling regime $g_\text{VO} > [\gamma_0, \kappa]$ \cite{Mi_Science_2016}. By fitting $A(f)/A_0$ at $\epsilon = 0$ [Fig.~\ref{fig:1}(e), right panel] and the results of microwave spectroscopy \cite{Supplement} to cavity input-output theory \cite{PhysRevB.94.195305}, we obtain $E_\text{L} = 37.5$ $\mu$eV, $E_\text{R} = 38.3$ $\mu$eV, $t = 24.3$ $\mu$eV, and $t' = 11.2$ $\mu$eV. The comparable magnitudes of $t$, $t'$, $E_\text{L}$ and $E_\text{R}$ have an important implication: the DQD energy levels are strongly influenced by hybridization of the valley and orbital d.o.f. 

The valley-orbit nature of DQD charge states is more clearly visualized through Landau-Zener-St\"uckelberg-Majorana (LZSM) interferometry \cite{Landau_1932,Zener_1932,Stuckelberg_1932,Majorana_1932,Oliver_Science_2005,Stehlik_PRB_2012,Forster_PRL_2017}. LZSM interferometry is performed by periodically driving the detuning parameter $\epsilon$, which in the time-domain sweeps the system through avoided crossings in the energy level diagram. The St\"uckelberg phase accumulated between avoided crossing traversals leads to a quantum interference pattern -- a ``fingerprint" that sensitively depends on the system's Hamiltonian \cite{Landau_1932,Zener_1932,Stuckelberg_1932,Majorana_1932,Shevchenko_Review_2010}.

We probe the energy level structure of the Si DQD in the time-domain by varying $\epsilon$ sinusoidally in time such that $\epsilon (t) = \epsilon_0 + eV_\text{ac} \cos \left( 2 \pi f_\text{g} t \right)$, where $f_\text{g}$ denotes the frequency of the drive and $t$ in this expression denotes time. We set $V_\text{B2} = 410$~mV for these measurements (resulting in $t = 25.4$ $\mu$eV and $t' = 11.8$ $\mu$eV), such that the device is in the dispersive regime with $E_\text{VO} / h - f_\text{c} \geq 55$ MHz $\gg g_\text{VO} / 2 \pi$ \cite{Mi_Science_2016,Mi_ValleyPRL_2017}. With these values of $t$ and $t'$ the cavity-induced Purcell decay rate is estimated to be $\Gamma_\text{P} / 2 \pi < 0.1$ MHz $\ll$ $\gamma_0 / 2 \pi$ such that charge decoherence is dominated by internal noise in the device, not coupling to the cavity mode \cite{Bienfait_Nature_2016}. In the dispersive regime, a change in the DQD charge state population leads to dispersive shift in the cavity transmission $A/A_0$ \cite{Stehlik_PRX, Koski_Arxiv_2018}. 

The LZSM interference pattern for this device is shown in Fig.~\ref{fig:2}(a), where we plot the steady-state cavity transmission amplitude $A/A_0$ as a function of $\epsilon_0$ and $eV_\text{ac}$ with $f_{\rm g}$ = 50 MHz. As  $eV_\text{ac}$ is increased from zero, we observe $>$20 interference fringes in a V-shaped region bounded by $eV_\text{ac} \approx |\epsilon_0|$. Within each interference fringe, a series of minima with $A/A_0 < 1$ are observed, indicating changes in the time-averaged population of the DQD charge states due to the evolution of the St\"uckelberg phase \cite{Oliver_Science_2005,Stehlik_PRB_2012,Forster_PRL_2017}. No interference fringes are observed outside of the V-shaped region, as here the detuning parameter is no longer swept through the avoided crossings near $\epsilon = 0$ [Fig.~\ref{fig:2}(b)].

The overall LZSM interference pattern observed in our device significantly deviates from previous work on superconducting and semiconductor qubits, where the interference fringes have an arc-like shape that are symmetric with respect to $\epsilon$ = 0 \cite{Oliver_Science_2005,Stehlik_PRB_2012,Forster_PRL_2017}. Instead, the interference fringes have the symmetric arc-like structure for $eV_\text{ac} < 60$ $\mu$eV, but  become increasingly more asymmetric at larger $eV_\text{ac}$ and resemble a harp. In addition, the concavity of the interference fringes changes from concave down to concave up around  $eV_\text{ac}$ $\approx$ 75 $\mu$eV. Finally, we observe clear quantum interference fringes with $f_{\rm g}$ = 50 MHz, which is roughly 200 times slower than GaAs charge qubit driving frequencies \cite{Stehlik_PRB_2012}.

The LZSM interference pattern may be qualitatively understood by considering the DQD energy level diagram [Fig.~\ref{fig:2}(b)]. For $\left| \epsilon \right| \gg E_\text{L,R}$, the DQD eigenstates are the unhybridized charge states $\ket{L}$, $\ket{L'}$, $\ket{R}$, $\ket{R'}$. For smaller values of $\epsilon$, the charge and valley d.o.f. hybridize through the tunnel couplings $t$ and $t'$ to form valley-orbit states. The quantum transition between the two lowest-lying energy states has a frequency $E_\text{VO} / h$, which is plotted in Fig.~\ref{fig:2}(c). For $\left| \epsilon \right| \leq E_\text{L,R}$, $E_\text{VO} / h$ has a quadratic dispersion relation as found in conventional two-level charge qubits \cite{Stehlik_PRB_2012,Forster_PRL_2017}. In contrast, for $\left| \epsilon \right| \gg E_\text{L,R}$, $E_\text{VO}$ approaches $E_\text{L}$ (for $\epsilon < 0$) or $E_\text{R}$ (for $\epsilon > 0$). Driving the DQD with $eV_\text{ac} \leq E_\text{L,R}$ therefore results in the arc-like interference patterns often associated with driven two-level systems \cite{Shevchenko_Review_2010}. However, once $eV_\text{ac} \gg E_\text{L,R}$ the energy difference between the ground state and first excited state is primarily set by single-dot valley splittings that are different for both dots, leading to an asymmetric interference pattern and a change in the concavity of the interference fringes.

To quantitatively compare the data to theory, we employ input-output theory, which provides the cavity response $A/A_0$ as a function of the DQD susceptibility $\chi$ \cite{Stehlik_PRX}. If the DQD is driven, a proper time-average of $\chi$ is required and can be derived within Floquet theory \cite{Kohler_PRL_2017}. At the low drive frequency $f_\text{g}$ used in this experiment, the Floquet states are approximated by adiabatic solutions of the Schr\"odinger equation \cite{Supplement}. Theoretical predictions for $A/A_0$ are shown in the inset of Fig.~\ref{fig:2}(a). The excellent agreement between experiment and theory confirms the energy level structure of the DQD charge states. An alternative interpretation of the LZSM interference pattern based on dressed states is given in \cite{Supplement}.

The data in Fig.~\ref{fig:2}(a) also yield information on the quantum coherence of the two lowest valley-orbit states away from $\epsilon$ = 0. For constructive interference of the St\"uckelberg phase, consecutive passages through the avoided crossing must occur within the coherence time of the system. As such, LZSM interferences are observable only if the time-averaged decoherence rate $\bar{\gamma} = \frac{1}{T} \int^T_0 \gamma \left[ \epsilon (t) \right] dt$ satisfies $\bar{\gamma} \lesssim f_\text{g}$ where $T = 1/f_\text{g}$ is the period of the drive and $t$ here denotes time \cite{Shevchenko_Review_2010,Forster_PRL_2017}. For typical semiconductor charge qubits, charge dephasing rates $\gamma_\phi$ are several GHz at $\epsilon \neq 0$ and LZSM interferometry must be performed at high drive frequencies $f_\text{g} \geq 2.5$ GHz to see an interference pattern \cite{Stehlik_PRB_2012,Forster_PRL_2017}. In contrast, we observe a clear quantum interference pattern with $f_\text{g} = 50$ MHz, which indicates long-lived charge coherence even when far detuned from $\epsilon$ = 0.

A primary factor contributing to the long coherence times of the valley-orbit states is evident from the relatively flat dispersion relation $E_\text{VO} (\epsilon)$. Based on the qubit parameters, $\left| dE_\text{VO}/d\epsilon \right|$ has a maximum value of 0.08 at $\epsilon = 30$ $\mu$eV and asymptotes to zero at large $\left| \epsilon \right|$. In contrast, the dispersion relation of a conventional charge qubit $E_\text{CQ} (\epsilon)$ yields $\left| dE_\text{CQ}/d\epsilon \right| \approx 1$ at finite $\epsilon$ [Fig.~\ref{fig:2}(c)].  Here and in Fig.~4(b), we have assumed $t_c$ = 25.4 $\mu$eV for the charge qubit such its minimum energy splitting is the same as the valley-orbit qubit. Based on the energy-level structure, charge-noise-induced fluctuations in $\epsilon$ will lead to markedly smaller fluctuations in the energy splitting between the valley-orbit states and lower decoherence rates \cite{Ithier-PRB_2005,Petersson_PRL_2010,Dial_Charge_2013,Forster_PRL_2017}. 

\begin{figure}[t]
	\centering
	\includegraphics[width=\columnwidth]{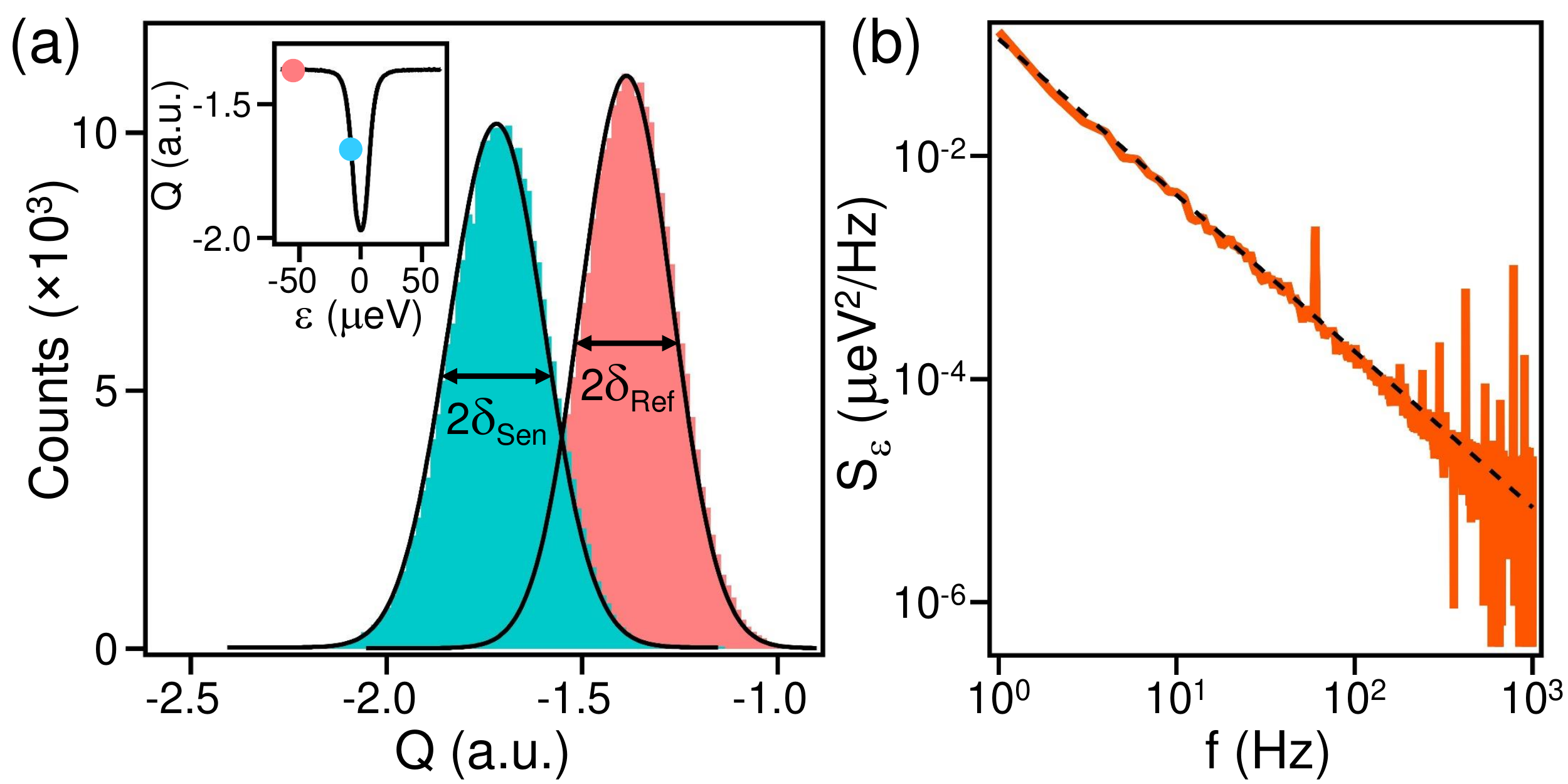}
	\caption{(a) Histograms showing the out-of-phase component of the cavity output field, $Q$, sampled at a frequency of 250 kHz over 1 sec.\ at $\epsilon = -8$ $\mu$eV (cyan) and $\epsilon = -60$ $\mu$eV (red). Black lines are fits to Gaussian functions. Inset shows $Q(\epsilon)$ and the colored dots represent the locations where the histograms are taken. $eV_{\rm ac}$ = 0 for these measurements. (b) Power spectral density of the noise in $\epsilon$, $S_\epsilon(f)$, with a fit to $S_\epsilon (f) \approx S_0[(1 \text{ Hz})/f]^{\beta}$. $S_\epsilon (f)$ is too small to be resolved beyond $f$ = 1 kHz using this method.}
	\label{fig:3}
\end{figure}

To evaluate whether $\left| dE_\text{VO}/d\epsilon \right|$ is sufficient to support the observed level of charge coherence, we explicitly measure the detuning noise \cite{Walraff_ChargeNoise_APL}. The out-of-phase component of the cavity output field, $Q$, is first recorded as a time-series at $\epsilon = -8$ $\mu$eV where $Q$ is strongly dependent on $\epsilon$ with a sensitivity $\left| dQ/d \epsilon \right| = C$ [inset to Fig.~\ref{fig:3}(a)]. The data, shown as a histogram in Fig.~\ref{fig:3}(a), have a Gaussian profile with standard deviation $\delta_\text{Sens}$. To separate the noise in $Q$ due to fluctuations in $\epsilon$ from background noise in the measurement setup, we also sample $Q$ at $\epsilon = -60$ $\mu$eV, where $dQ/d \epsilon \approx 0$. These data [Fig.~\ref{fig:3}(a)] have a standard deviation $\delta_\text{Ref}$. The standard deviation of the detuning fluctuations is then $\delta_\epsilon = \frac{1}{C} \sqrt{\delta_\text{Sens}^2 - \delta_\text{Ref}^2}$ = 0.87 $\mu$eV. Factoring in the slope of the dispersion relation, this detuning fluctuation corresponds to a maximum fluctuation in  $E_\text{VO}/h$ = $\left| dE_\text{VO}/d\epsilon \right|_{\rm max}$  $\times$ 0.87 $\mu$eV $\approx 17$ MHz. Our simple analysis shows that the magnitude of the charge noise, combined with the DQD energy level structure, is sufficient to explain the observation of a LZSM interference pattern at such low drive frequencies. Looking forward, it may be helpful to more precisely model the effects of charge noise using simulated lever-arms and the charge noise insensitivity parameter \cite{Reed_PRL_2016}.

Discrete Fourier transforms of the time-series allow the power spectral density of the detuning noise $S_\epsilon (f)$ to be determined [Fig.~\ref{fig:3}(b)]. At low frequencies the power spectrum scales as $S_\epsilon (f) \approx S_0[(1 \text{ Hz})/f]^{\beta}$ with $S_0$ = 0.11 $\mu\text{eV}^2/{\text{Hz}}$ and $\beta$ $\approx$ 1.4. Using this noise spectrum we calculate a maximum dephasing rate $\gamma_\phi / 2 \pi \approx 6$ MHz at $\left| \epsilon \right| = 30$ $\mu$eV, which is indeed below the drive frequency $f_\text{g} = 50$ MHz. Converting to units of electron charge, we find $S_\text{c} = (e/E_\text{c})^2 S_\epsilon$ = $3.8 \times 10^{-9}$ $e^2$/Hz at a $f$ = 1 Hz, where $E_\text{c} \approx 5.4$ meV is the charging energy \cite{Supplement}.

The valley-orbit states may serve as the basis states of a highly controllable and coherent hybrid qubit. First, the orbital nature of the hybrid qubit allows fast manipulation using microwaves, as evidenced by its strong-coupling to a single photon in the cavity [Fig.~\ref{fig:1}(e)]. Second, unlike conventional two-level charge qubits, the valley-orbit qubit may be operated away from $\epsilon$ = 0 without significant loss of coherence due to its relatively flat energy bands and small charge noise sensitivity \cite{Koch_PRA_2007}. 

\begin{figure}[t]
	\centering
	\includegraphics[width=\columnwidth]{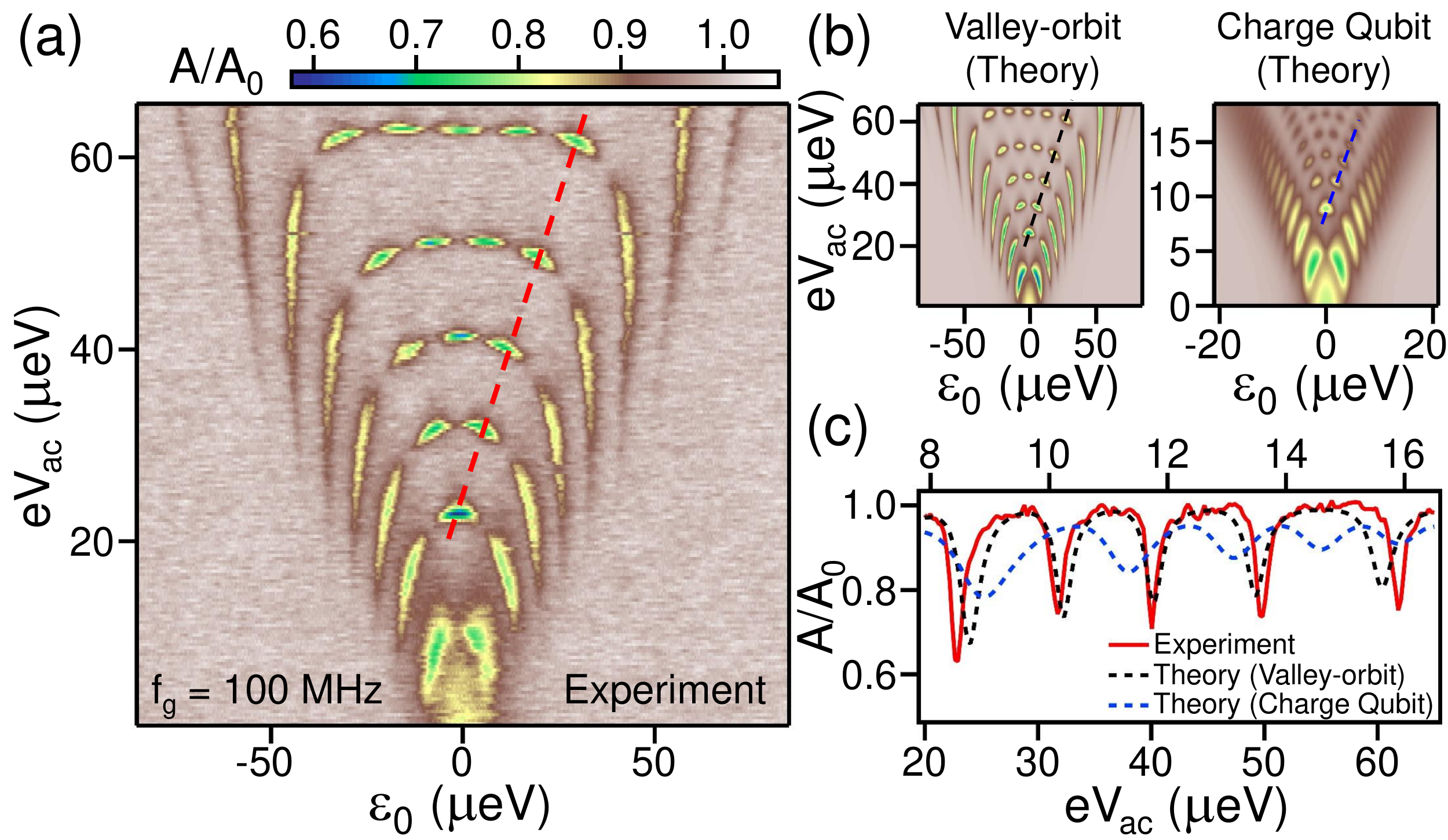}
	\caption{(a) $A/A_0$ as a function of $\epsilon_0$ and $eV_\text{ac}$, with $f_\text{g} = 100$ MHz. (b) Theoretically predicted LZSM interference patterns obtained from a 4-level valley-orbit model (left) and a standard two-level charge qubit model (right). The color-scale is the same as in panel (a). (c) $A/A_0$ as a function of $eV_\text{ac}$, taken along the dashed lines in panels (a) and (b). The experimental data and calculation based on valley-orbit states are plotted along the bottom $x$-axis, whereas the calculation based on a charge qubit is plotted along the top $x$-axis.}
	\label{fig:4}
\end{figure}

As a demonstration, we measure $A (\epsilon_0, eV_\text{ac})/A_0$ with $f_\text{g} = 100$ MHz [Fig.~\ref{fig:4}(a)]. Here we observe clearly resolved interference minima that show little decay in intensity as the drive amplitude $eV_\text{ac}$ increases, further confirming that the decoherence rate of the valley-states does not increase appreciably at large DQD detunings. To compare with theoretical expectations, we calculate $A (\epsilon_0, eV_\text{ac})/A_0$ using experimental parameters [see the left panel of Fig.~\ref{fig:4}(b)] and find excellent agreement with the data. In comparison, the interference patterns calculated for a conventional charge qubit [right panel of Fig.~\ref{fig:4}(b)] decay rapidly as $eV_\text{ac}$ increases due to fast decoherence at large DQD detunings. These simulations are more directly compared by plotting $A/A_0$ along a line connecting the center of a resonance minimum within each fringe [Fig.~\ref{fig:4}(c)].

In conclusion, we observe LZSM interference patterns in a cavity-coupled Si DQD charge qubit when the DQD detuning is driven at frequencies as low as 50 MHz. Analysis of the interference patterns reveals that the basis states for the DQD charge qubit are hybridized valley-orbit states rather than pure orbital states. Compared to conventional charge qubits, and other recently developed hybrid qubits that all rely on operation at sweet spots to maintain coherence \cite{Petersson_PRL_2010,Kim_Hybrid_2014,Schoenfield_NatComm_2017}, qubits formed by the valley-orbit states have a small sensitivity to charge noise, even at arbitrary detunings. Analogous to superconducting transmon qubits, the charge noise insensitivity of the valley-orbit states arises from their flat energy level structure \cite{Koch_PRA_2007}. Deterministic control of valley splittings, perhaps made possible through further material research efforts, has the potential of turning the valley d.o.f.\ in Si into a powerful resource for reducing the charge noise sensitivities of silicon qubits.

\begin{acknowledgments}
We acknowledge valuable discussions with M.~J.~Gullans, J.~Kerckhoff, T.~D.~Ladd, F.~Nori, M.~S.~Rudner, S.~N.~Shevchenko and J.~M.~Taylor. Supported by the U.S. Department of Defense under contract H98230-15-C0453, Army Research Office grant W911NF-15-1-0149, the Gordon and Betty Moore Foundations EPiQS Initiative through grant GBMF4535, and by the Spanish Ministry of Economy and Competiveness through Grant MAT2017-86717-P. Devices were fabricated in the Princeton University Quantum Device Nanofabrication Laboratory.
\end{acknowledgments}

\bibliographystyle{apsrev4-1}
\bibliography{references_v11}

\begin{thebibliography}{36}%
\makeatletter
\providecommand \@ifxundefined [1]{%
 \@ifx{#1\undefined}
}%
\providecommand \@ifnum [1]{%
 \ifnum #1\expandafter \@firstoftwo
 \else \expandafter \@secondoftwo
 \fi
}%
\providecommand \@ifx [1]{%
 \ifx #1\expandafter \@firstoftwo
 \else \expandafter \@secondoftwo
 \fi
}%
\providecommand \natexlab [1]{#1}%
\providecommand \enquote  [1]{``#1''}%
\providecommand \bibnamefont  [1]{#1}%
\providecommand \bibfnamefont [1]{#1}%
\providecommand \citenamefont [1]{#1}%
\providecommand \href@noop [0]{\@secondoftwo}%
\providecommand \href [0]{\begingroup \@sanitize@url \@href}%
\providecommand \@href[1]{\@@startlink{#1}\@@href}%
\providecommand \@@href[1]{\endgroup#1\@@endlink}%
\providecommand \@sanitize@url [0]{\catcode `\\12\catcode `\$12\catcode
  `\&12\catcode `\#12\catcode `\^12\catcode `\_12\catcode `\%12\relax}%
\providecommand \@@startlink[1]{}%
\providecommand \@@endlink[0]{}%
\providecommand \url  [0]{\begingroup\@sanitize@url \@url }%
\providecommand \@url [1]{\endgroup\@href {#1}{\urlprefix }}%
\providecommand \urlprefix  [0]{URL }%
\providecommand \Eprint [0]{\href }%
\providecommand \doibase [0]{http://dx.doi.org/}%
\providecommand \selectlanguage [0]{\@gobble}%
\providecommand \bibinfo  [0]{\@secondoftwo}%
\providecommand \bibfield  [0]{\@secondoftwo}%
\providecommand \translation [1]{[#1]}%
\providecommand \BibitemOpen [0]{}%
\providecommand \bibitemStop [0]{}%
\providecommand \bibitemNoStop [0]{.\EOS\space}%
\providecommand \EOS [0]{\spacefactor3000\relax}%
\providecommand \BibitemShut  [1]{\csname bibitem#1\endcsname}%
\let\auto@bib@innerbib\@empty
\bibitem [{\citenamefont {Lestas}\ \emph {et~al.}(2010)\citenamefont {Lestas},
  \citenamefont {Vinnicombe},\ and\ \citenamefont
  {Paulsson}}]{Lestas_Nature_2010}%
  \BibitemOpen
  \bibfield  {author} {\bibinfo {author} {\bibfnamefont {I.}~\bibnamefont
  {Lestas}}, \bibinfo {author} {\bibfnamefont {G.}~\bibnamefont {Vinnicombe}},
  \ and\ \bibinfo {author} {\bibfnamefont {J.}~\bibnamefont {Paulsson}},\
  }\href@noop {} {\bibfield  {journal} {\bibinfo  {journal} {Nature (London)}\
  }\textbf {\bibinfo {volume} {467}},\ \bibinfo {pages} {174} (\bibinfo {year}
  {2010})}\BibitemShut {NoStop}%
\bibitem [{\citenamefont {Weissman}(1988)}]{Weissman_RMP_1988}%
  \BibitemOpen
  \bibfield  {author} {\bibinfo {author} {\bibfnamefont {M.~B.}\ \bibnamefont
  {Weissman}},\ }\href@noop {} {\bibfield  {journal} {\bibinfo  {journal} {Rev.
  Mod. Phys.}\ }\textbf {\bibinfo {volume} {60}},\ \bibinfo {pages} {537}
  (\bibinfo {year} {1988})}\BibitemShut {NoStop}%
\bibitem [{\citenamefont {Paladino}\ \emph {et~al.}(2014)\citenamefont
  {Paladino}, \citenamefont {Galperin}, \citenamefont {Falci},\ and\
  \citenamefont {Altshuler}}]{Paladino_RMP_2014}%
  \BibitemOpen
  \bibfield  {author} {\bibinfo {author} {\bibfnamefont {E.}~\bibnamefont
  {Paladino}}, \bibinfo {author} {\bibfnamefont {Y.~M.}\ \bibnamefont
  {Galperin}}, \bibinfo {author} {\bibfnamefont {G.}~\bibnamefont {Falci}}, \
  and\ \bibinfo {author} {\bibfnamefont {B.~L.}\ \bibnamefont {Altshuler}},\
  }\href@noop {} {\bibfield  {journal} {\bibinfo  {journal} {Rev. Mod. Phys.}\
  }\textbf {\bibinfo {volume} {86}},\ \bibinfo {pages} {361} (\bibinfo {year}
  {2014})}\BibitemShut {NoStop}%
\bibitem [{\citenamefont {Nakamura}\ \emph {et~al.}(1999)\citenamefont
  {Nakamura}, \citenamefont {Pashkin},\ and\ \citenamefont
  {Tsai}}]{Nakamura_Nature_1999}%
  \BibitemOpen
  \bibfield  {author} {\bibinfo {author} {\bibfnamefont {Y.}~\bibnamefont
  {Nakamura}}, \bibinfo {author} {\bibfnamefont {Y.~A.}\ \bibnamefont
  {Pashkin}}, \ and\ \bibinfo {author} {\bibfnamefont {J.~S.}\ \bibnamefont
  {Tsai}},\ }\href@noop {} {\bibfield  {journal} {\bibinfo  {journal} {Nature
  (London)}\ }\textbf {\bibinfo {volume} {398}},\ \bibinfo {pages} {786}
  (\bibinfo {year} {1999})}\BibitemShut {NoStop}%
\bibitem [{\citenamefont {Ithier}\ \emph {et~al.}(2005)\citenamefont {Ithier},
  \citenamefont {Collin}, \citenamefont {Joyez}, \citenamefont {Meeson},
  \citenamefont {Vion}, \citenamefont {Esteve}, \citenamefont {Chiarello},
  \citenamefont {Shnirman}, \citenamefont {Makhlin}, \citenamefont {Schriefl},\
  and\ \citenamefont {Sch\"on}}]{Ithier-PRB_2005}%
  \BibitemOpen
  \bibfield  {author} {\bibinfo {author} {\bibfnamefont {G.}~\bibnamefont
  {Ithier}}, \bibinfo {author} {\bibfnamefont {E.}~\bibnamefont {Collin}},
  \bibinfo {author} {\bibfnamefont {P.}~\bibnamefont {Joyez}}, \bibinfo
  {author} {\bibfnamefont {P.~J.}\ \bibnamefont {Meeson}}, \bibinfo {author}
  {\bibfnamefont {D.}~\bibnamefont {Vion}}, \bibinfo {author} {\bibfnamefont
  {D.}~\bibnamefont {Esteve}}, \bibinfo {author} {\bibfnamefont
  {F.}~\bibnamefont {Chiarello}}, \bibinfo {author} {\bibfnamefont
  {A.}~\bibnamefont {Shnirman}}, \bibinfo {author} {\bibfnamefont
  {Y.}~\bibnamefont {Makhlin}}, \bibinfo {author} {\bibfnamefont
  {J.}~\bibnamefont {Schriefl}}, \ and\ \bibinfo {author} {\bibfnamefont
  {G.}~\bibnamefont {Sch\"on}},\ }\href@noop {} {\bibfield  {journal} {\bibinfo
   {journal} {Phys. Rev. B}\ }\textbf {\bibinfo {volume} {72}},\ \bibinfo
  {pages} {134519} (\bibinfo {year} {2005})}\BibitemShut {NoStop}%
\bibitem [{\citenamefont {Yoneda}\ \emph {et~al.}(2018)\citenamefont {Yoneda},
  \citenamefont {Takeda}, \citenamefont {Otsuka}, \citenamefont {Nakajima},
  \citenamefont {Delbecq}, \citenamefont {Allison}, \citenamefont {Honda},
  \citenamefont {Kodera}, \citenamefont {Oda}, \citenamefont {Hoshi},
  \citenamefont {Usami}, \citenamefont {Itoh},\ and\ \citenamefont
  {Tarucha}}]{Yoneda_threenines_2017}%
  \BibitemOpen
  \bibfield  {author} {\bibinfo {author} {\bibfnamefont {J.}~\bibnamefont
  {Yoneda}}, \bibinfo {author} {\bibfnamefont {K.}~\bibnamefont {Takeda}},
  \bibinfo {author} {\bibfnamefont {T.}~\bibnamefont {Otsuka}}, \bibinfo
  {author} {\bibfnamefont {T.}~\bibnamefont {Nakajima}}, \bibinfo {author}
  {\bibfnamefont {M.~R.}\ \bibnamefont {Delbecq}}, \bibinfo {author}
  {\bibfnamefont {G.}~\bibnamefont {Allison}}, \bibinfo {author} {\bibfnamefont
  {T.}~\bibnamefont {Honda}}, \bibinfo {author} {\bibfnamefont
  {T.}~\bibnamefont {Kodera}}, \bibinfo {author} {\bibfnamefont
  {S.}~\bibnamefont {Oda}}, \bibinfo {author} {\bibfnamefont {Y.}~\bibnamefont
  {Hoshi}}, \bibinfo {author} {\bibfnamefont {N.}~\bibnamefont {Usami}},
  \bibinfo {author} {\bibfnamefont {K.~M.}\ \bibnamefont {Itoh}}, \ and\
  \bibinfo {author} {\bibfnamefont {S.}~\bibnamefont {Tarucha}},\ }\href@noop
  {} {\bibfield  {journal} {\bibinfo  {journal} {Nat. Nanotechnol.}\ }\textbf
  {\bibinfo {volume} {13}},\ \bibinfo {pages} {102} (\bibinfo {year}
  {2018})}\BibitemShut {NoStop}%
\bibitem [{\citenamefont {Petta}\ \emph {et~al.}(2005)\citenamefont {Petta},
  \citenamefont {Johnson}, \citenamefont {Taylor}, \citenamefont {Laird},
  \citenamefont {Yacoby}, \citenamefont {Lukin}, \citenamefont {Marcus},
  \citenamefont {Hanson},\ and\ \citenamefont {Gossard}}]{Petta_Science}%
  \BibitemOpen
  \bibfield  {author} {\bibinfo {author} {\bibfnamefont {J.~R.}\ \bibnamefont
  {Petta}}, \bibinfo {author} {\bibfnamefont {A.~C.}\ \bibnamefont {Johnson}},
  \bibinfo {author} {\bibfnamefont {J.~M.}\ \bibnamefont {Taylor}}, \bibinfo
  {author} {\bibfnamefont {E.~A.}\ \bibnamefont {Laird}}, \bibinfo {author}
  {\bibfnamefont {A.}~\bibnamefont {Yacoby}}, \bibinfo {author} {\bibfnamefont
  {M.~D.}\ \bibnamefont {Lukin}}, \bibinfo {author} {\bibfnamefont {C.~M.}\
  \bibnamefont {Marcus}}, \bibinfo {author} {\bibfnamefont {M.~P.}\
  \bibnamefont {Hanson}}, \ and\ \bibinfo {author} {\bibfnamefont {A.~C.}\
  \bibnamefont {Gossard}},\ }\href@noop {} {\bibfield  {journal} {\bibinfo
  {journal} {Science}\ }\textbf {\bibinfo {volume} {309}},\ \bibinfo {pages}
  {2180} (\bibinfo {year} {2005})}\BibitemShut {NoStop}%
\bibitem [{\citenamefont {Dial}\ \emph {et~al.}(2013)\citenamefont {Dial},
  \citenamefont {Shulman}, \citenamefont {Harvey}, \citenamefont {Bluhm},
  \citenamefont {Umansky},\ and\ \citenamefont {Yacoby}}]{Dial_Charge_2013}%
  \BibitemOpen
  \bibfield  {author} {\bibinfo {author} {\bibfnamefont {O.~E.}\ \bibnamefont
  {Dial}}, \bibinfo {author} {\bibfnamefont {M.~D.}\ \bibnamefont {Shulman}},
  \bibinfo {author} {\bibfnamefont {S.~P.}\ \bibnamefont {Harvey}}, \bibinfo
  {author} {\bibfnamefont {H.}~\bibnamefont {Bluhm}}, \bibinfo {author}
  {\bibfnamefont {V.}~\bibnamefont {Umansky}}, \ and\ \bibinfo {author}
  {\bibfnamefont {A.}~\bibnamefont {Yacoby}},\ }\href@noop {} {\bibfield
  {journal} {\bibinfo  {journal} {Phys. Rev. Lett.}\ }\textbf {\bibinfo
  {volume} {110}},\ \bibinfo {pages} {146804} (\bibinfo {year}
  {2013})}\BibitemShut {NoStop}%
\bibitem [{\citenamefont {Vion}\ \emph {et~al.}(2002)\citenamefont {Vion},
  \citenamefont {Aassime}, \citenamefont {Cottet}, \citenamefont {Joyez},
  \citenamefont {Pothier}, \citenamefont {Urbina}, \citenamefont {Esteve},\
  and\ \citenamefont {Devoret}}]{Vion_Science_2002}%
  \BibitemOpen
  \bibfield  {author} {\bibinfo {author} {\bibfnamefont {D.}~\bibnamefont
  {Vion}}, \bibinfo {author} {\bibfnamefont {A.}~\bibnamefont {Aassime}},
  \bibinfo {author} {\bibfnamefont {A.}~\bibnamefont {Cottet}}, \bibinfo
  {author} {\bibfnamefont {P.}~\bibnamefont {Joyez}}, \bibinfo {author}
  {\bibfnamefont {H.}~\bibnamefont {Pothier}}, \bibinfo {author} {\bibfnamefont
  {C.}~\bibnamefont {Urbina}}, \bibinfo {author} {\bibfnamefont
  {D.}~\bibnamefont {Esteve}}, \ and\ \bibinfo {author} {\bibfnamefont {M.~H.}\
  \bibnamefont {Devoret}},\ }\href@noop {} {\bibfield  {journal} {\bibinfo
  {journal} {Science}\ }\textbf {\bibinfo {volume} {296}},\ \bibinfo {pages}
  {886} (\bibinfo {year} {2002})}\BibitemShut {NoStop}%
\bibitem [{\citenamefont {Schreier}\ \emph {et~al.}(2008)\citenamefont
  {Schreier}, \citenamefont {Houck}, \citenamefont {Koch}, \citenamefont
  {Schuster}, \citenamefont {Johnson}, \citenamefont {Chow}, \citenamefont
  {Gambetta}, \citenamefont {Majer}, \citenamefont {Frunzio}, \citenamefont
  {Devoret}, \citenamefont {Girvin},\ and\ \citenamefont
  {Schoelkopf}}]{Schrier_PRB_2008}%
  \BibitemOpen
  \bibfield  {author} {\bibinfo {author} {\bibfnamefont {J.~A.}\ \bibnamefont
  {Schreier}}, \bibinfo {author} {\bibfnamefont {A.~A.}\ \bibnamefont {Houck}},
  \bibinfo {author} {\bibfnamefont {J.}~\bibnamefont {Koch}}, \bibinfo {author}
  {\bibfnamefont {D.~I.}\ \bibnamefont {Schuster}}, \bibinfo {author}
  {\bibfnamefont {B.~R.}\ \bibnamefont {Johnson}}, \bibinfo {author}
  {\bibfnamefont {J.~M.}\ \bibnamefont {Chow}}, \bibinfo {author}
  {\bibfnamefont {J.~M.}\ \bibnamefont {Gambetta}}, \bibinfo {author}
  {\bibfnamefont {J.}~\bibnamefont {Majer}}, \bibinfo {author} {\bibfnamefont
  {L.}~\bibnamefont {Frunzio}}, \bibinfo {author} {\bibfnamefont {M.~H.}\
  \bibnamefont {Devoret}}, \bibinfo {author} {\bibfnamefont {S.~M.}\
  \bibnamefont {Girvin}}, \ and\ \bibinfo {author} {\bibfnamefont {R.~J.}\
  \bibnamefont {Schoelkopf}},\ }\href@noop {} {\bibfield  {journal} {\bibinfo
  {journal} {Phys. Rev. B}\ }\textbf {\bibinfo {volume} {77}},\ \bibinfo
  {pages} {080502} (\bibinfo {year} {2008})}\BibitemShut {NoStop}%
\bibitem [{\citenamefont {Petersson}\ \emph {et~al.}(2010)\citenamefont
  {Petersson}, \citenamefont {Petta}, \citenamefont {Lu},\ and\ \citenamefont
  {Gossard}}]{Petersson_PRL_2010}%
  \BibitemOpen
  \bibfield  {author} {\bibinfo {author} {\bibfnamefont {K.~D.}\ \bibnamefont
  {Petersson}}, \bibinfo {author} {\bibfnamefont {J.~R.}\ \bibnamefont
  {Petta}}, \bibinfo {author} {\bibfnamefont {H.}~\bibnamefont {Lu}}, \ and\
  \bibinfo {author} {\bibfnamefont {A.~C.}\ \bibnamefont {Gossard}},\
  }\href@noop {} {\bibfield  {journal} {\bibinfo  {journal} {Phys. Rev. Lett.}\
  }\textbf {\bibinfo {volume} {105}},\ \bibinfo {pages} {246804} (\bibinfo
  {year} {2010})}\BibitemShut {NoStop}%
\bibitem [{\citenamefont {Reed}\ \emph {et~al.}(2016)\citenamefont {Reed},
  \citenamefont {Maune}, \citenamefont {Andrews}, \citenamefont {Borselli},
  \citenamefont {Eng}, \citenamefont {Jura}, \citenamefont {Kiselev},
  \citenamefont {Ladd}, \citenamefont {Merkel}, \citenamefont {Milosavljevic},
  \citenamefont {Pritchett}, \citenamefont {Rakher}, \citenamefont {Ross},
  \citenamefont {Schmitz}, \citenamefont {Smith}, \citenamefont {Wright},
  \citenamefont {Gyure},\ and\ \citenamefont {Hunter}}]{Reed_PRL_2016}%
  \BibitemOpen
  \bibfield  {author} {\bibinfo {author} {\bibfnamefont {M.~D.}\ \bibnamefont
  {Reed}}, \bibinfo {author} {\bibfnamefont {B.~M.}\ \bibnamefont {Maune}},
  \bibinfo {author} {\bibfnamefont {R.~W.}\ \bibnamefont {Andrews}}, \bibinfo
  {author} {\bibfnamefont {M.~G.}\ \bibnamefont {Borselli}}, \bibinfo {author}
  {\bibfnamefont {K.}~\bibnamefont {Eng}}, \bibinfo {author} {\bibfnamefont
  {M.~P.}\ \bibnamefont {Jura}}, \bibinfo {author} {\bibfnamefont {A.~A.}\
  \bibnamefont {Kiselev}}, \bibinfo {author} {\bibfnamefont {T.~D.}\
  \bibnamefont {Ladd}}, \bibinfo {author} {\bibfnamefont {S.~T.}\ \bibnamefont
  {Merkel}}, \bibinfo {author} {\bibfnamefont {I.}~\bibnamefont
  {Milosavljevic}}, \bibinfo {author} {\bibfnamefont {E.~J.}\ \bibnamefont
  {Pritchett}}, \bibinfo {author} {\bibfnamefont {M.~T.}\ \bibnamefont
  {Rakher}}, \bibinfo {author} {\bibfnamefont {R.~S.}\ \bibnamefont {Ross}},
  \bibinfo {author} {\bibfnamefont {A.~E.}\ \bibnamefont {Schmitz}}, \bibinfo
  {author} {\bibfnamefont {A.}~\bibnamefont {Smith}}, \bibinfo {author}
  {\bibfnamefont {J.~A.}\ \bibnamefont {Wright}}, \bibinfo {author}
  {\bibfnamefont {M.~F.}\ \bibnamefont {Gyure}}, \ and\ \bibinfo {author}
  {\bibfnamefont {A.~T.}\ \bibnamefont {Hunter}},\ }\href@noop {} {\bibfield
  {journal} {\bibinfo  {journal} {Phys. Rev. Lett.}\ }\textbf {\bibinfo
  {volume} {116}},\ \bibinfo {pages} {110402} (\bibinfo {year}
  {2016})}\BibitemShut {NoStop}%
\bibitem [{\citenamefont {Martins}\ \emph {et~al.}(2016)\citenamefont
  {Martins}, \citenamefont {Malinowski}, \citenamefont {Nissen}, \citenamefont
  {Barnes}, \citenamefont {Fallahi}, \citenamefont {Gardner}, \citenamefont
  {Manfra}, \citenamefont {Marcus},\ and\ \citenamefont
  {Kuemmeth}}]{Martins_PRL_2016}%
  \BibitemOpen
  \bibfield  {author} {\bibinfo {author} {\bibfnamefont {F.}~\bibnamefont
  {Martins}}, \bibinfo {author} {\bibfnamefont {F.~K.}\ \bibnamefont
  {Malinowski}}, \bibinfo {author} {\bibfnamefont {P.~D.}\ \bibnamefont
  {Nissen}}, \bibinfo {author} {\bibfnamefont {E.}~\bibnamefont {Barnes}},
  \bibinfo {author} {\bibfnamefont {S.}~\bibnamefont {Fallahi}}, \bibinfo
  {author} {\bibfnamefont {G.~C.}\ \bibnamefont {Gardner}}, \bibinfo {author}
  {\bibfnamefont {M.~J.}\ \bibnamefont {Manfra}}, \bibinfo {author}
  {\bibfnamefont {C.~M.}\ \bibnamefont {Marcus}}, \ and\ \bibinfo {author}
  {\bibfnamefont {F.}~\bibnamefont {Kuemmeth}},\ }\href@noop {} {\bibfield
  {journal} {\bibinfo  {journal} {Phys. Rev. Lett.}\ }\textbf {\bibinfo
  {volume} {116}},\ \bibinfo {pages} {116801} (\bibinfo {year}
  {2016})}\BibitemShut {NoStop}%
\bibitem [{\citenamefont {Kim}\ \emph {et~al.}(2014)\citenamefont {Kim},
  \citenamefont {Shi}, \citenamefont {Simmons}, \citenamefont {Ward},
  \citenamefont {Prance}, \citenamefont {Koh}, \citenamefont {Gamble},
  \citenamefont {Savage}, \citenamefont {Lagally}, \citenamefont {Friesen},
  \citenamefont {Coppersmith},\ and\ \citenamefont
  {Eriksson}}]{Kim_Hybrid_2014}%
  \BibitemOpen
  \bibfield  {author} {\bibinfo {author} {\bibfnamefont {D.}~\bibnamefont
  {Kim}}, \bibinfo {author} {\bibfnamefont {Z.}~\bibnamefont {Shi}}, \bibinfo
  {author} {\bibfnamefont {C.~B.}\ \bibnamefont {Simmons}}, \bibinfo {author}
  {\bibfnamefont {D.~R.}\ \bibnamefont {Ward}}, \bibinfo {author}
  {\bibfnamefont {J.~R.}\ \bibnamefont {Prance}}, \bibinfo {author}
  {\bibfnamefont {T.~S.}\ \bibnamefont {Koh}}, \bibinfo {author} {\bibfnamefont
  {J.~K.}\ \bibnamefont {Gamble}}, \bibinfo {author} {\bibfnamefont {D.~E.}\
  \bibnamefont {Savage}}, \bibinfo {author} {\bibfnamefont {M.~G.}\
  \bibnamefont {Lagally}}, \bibinfo {author} {\bibfnamefont {M.}~\bibnamefont
  {Friesen}}, \bibinfo {author} {\bibfnamefont {S.~N.}\ \bibnamefont
  {Coppersmith}}, \ and\ \bibinfo {author} {\bibfnamefont {M.~A.}\ \bibnamefont
  {Eriksson}},\ }\href@noop {} {\bibfield  {journal} {\bibinfo  {journal}
  {Nature (London)}\ }\textbf {\bibinfo {volume} {511}},\ \bibinfo {pages} {70}
  (\bibinfo {year} {2014})}\BibitemShut {NoStop}%
\bibitem [{\citenamefont {Sch\"affler}(1997)}]{Schaffler_SiGe_Review}%
  \BibitemOpen
  \bibfield  {author} {\bibinfo {author} {\bibfnamefont {F.}~\bibnamefont
  {Sch\"affler}},\ }\href@noop {} {\bibfield  {journal} {\bibinfo  {journal}
  {Semi. Sci. Technol.}\ }\textbf {\bibinfo {volume} {12}},\ \bibinfo {pages}
  {1515} (\bibinfo {year} {1997})}\BibitemShut {NoStop}%
\bibitem [{\citenamefont {Borselli}\ \emph {et~al.}(2011)\citenamefont
  {Borselli}, \citenamefont {Ross}, \citenamefont {Kiselev}, \citenamefont
  {Croke}, \citenamefont {Holabird}, \citenamefont {Deelman}, \citenamefont
  {Warren}, \citenamefont {Alvarado-Rodriguez}, \citenamefont {Milosavljevic},
  \citenamefont {Ku}, \citenamefont {Wong}, \citenamefont {Schmitz},
  \citenamefont {Sokolich}, \citenamefont {Gyure},\ and\ \citenamefont
  {Hunter}}]{Borselli_ValleyAPL_2011}%
  \BibitemOpen
  \bibfield  {author} {\bibinfo {author} {\bibfnamefont {M.~G.}\ \bibnamefont
  {Borselli}}, \bibinfo {author} {\bibfnamefont {R.~S.}\ \bibnamefont {Ross}},
  \bibinfo {author} {\bibfnamefont {A.~A.}\ \bibnamefont {Kiselev}}, \bibinfo
  {author} {\bibfnamefont {E.~T.}\ \bibnamefont {Croke}}, \bibinfo {author}
  {\bibfnamefont {K.~S.}\ \bibnamefont {Holabird}}, \bibinfo {author}
  {\bibfnamefont {P.~W.}\ \bibnamefont {Deelman}}, \bibinfo {author}
  {\bibfnamefont {L.~D.}\ \bibnamefont {Warren}}, \bibinfo {author}
  {\bibfnamefont {I.}~\bibnamefont {Alvarado-Rodriguez}}, \bibinfo {author}
  {\bibfnamefont {I.}~\bibnamefont {Milosavljevic}}, \bibinfo {author}
  {\bibfnamefont {F.~C.}\ \bibnamefont {Ku}}, \bibinfo {author} {\bibfnamefont
  {W.~S.}\ \bibnamefont {Wong}}, \bibinfo {author} {\bibfnamefont {A.~E.}\
  \bibnamefont {Schmitz}}, \bibinfo {author} {\bibfnamefont {M.}~\bibnamefont
  {Sokolich}}, \bibinfo {author} {\bibfnamefont {M.~F.}\ \bibnamefont {Gyure}},
  \ and\ \bibinfo {author} {\bibfnamefont {A.~T.}\ \bibnamefont {Hunter}},\
  }\href@noop {} {\bibfield  {journal} {\bibinfo  {journal} {Appl. Phys.
  Lett.}\ }\textbf {\bibinfo {volume} {98}},\ \bibinfo {eid} {123118} (\bibinfo
  {year} {2011})}\BibitemShut {NoStop}%
\bibitem [{\citenamefont {Mi}\ \emph {et~al.}(2017{\natexlab{a}})\citenamefont
  {Mi}, \citenamefont {P\'eterfalvi}, \citenamefont {Burkard},\ and\
  \citenamefont {Petta}}]{Mi_ValleyPRL_2017}%
  \BibitemOpen
  \bibfield  {author} {\bibinfo {author} {\bibfnamefont {X.}~\bibnamefont
  {Mi}}, \bibinfo {author} {\bibfnamefont {C.~G.}\ \bibnamefont
  {P\'eterfalvi}}, \bibinfo {author} {\bibfnamefont {G.}~\bibnamefont
  {Burkard}}, \ and\ \bibinfo {author} {\bibfnamefont {J.~R.}\ \bibnamefont
  {Petta}},\ }\href@noop {} {\bibfield  {journal} {\bibinfo  {journal} {Phys.
  Rev. Lett.}\ }\textbf {\bibinfo {volume} {119}},\ \bibinfo {pages} {176803}
  (\bibinfo {year} {2017}{\natexlab{a}})}\BibitemShut {NoStop}%
\bibitem [{\citenamefont {Schoenfield}\ \emph {et~al.}(2017)\citenamefont
  {Schoenfield}, \citenamefont {Freeman},\ and\ \citenamefont
  {Jiang}}]{Schoenfield_NatComm_2017}%
  \BibitemOpen
  \bibfield  {author} {\bibinfo {author} {\bibfnamefont {J.~S.}\ \bibnamefont
  {Schoenfield}}, \bibinfo {author} {\bibfnamefont {B.~M.}\ \bibnamefont
  {Freeman}}, \ and\ \bibinfo {author} {\bibfnamefont {H.}~\bibnamefont
  {Jiang}},\ }\href@noop {} {\bibfield  {journal} {\bibinfo  {journal} {Nat.
  Commun.}\ }\textbf {\bibinfo {volume} {8}},\ \bibinfo {pages} {64} (\bibinfo
  {year} {2017})}\BibitemShut {NoStop}%
\bibitem [{\citenamefont {Koch}\ \emph {et~al.}(2007)\citenamefont {Koch},
  \citenamefont {Yu}, \citenamefont {Gambetta}, \citenamefont {Houck},
  \citenamefont {Schuster}, \citenamefont {Majer}, \citenamefont {Blais},
  \citenamefont {Devoret}, \citenamefont {Girvin},\ and\ \citenamefont
  {Schoelkopf}}]{Koch_PRA_2007}%
  \BibitemOpen
  \bibfield  {author} {\bibinfo {author} {\bibfnamefont {J.}~\bibnamefont
  {Koch}}, \bibinfo {author} {\bibfnamefont {T.~M.}\ \bibnamefont {Yu}},
  \bibinfo {author} {\bibfnamefont {J.}~\bibnamefont {Gambetta}}, \bibinfo
  {author} {\bibfnamefont {A.~A.}\ \bibnamefont {Houck}}, \bibinfo {author}
  {\bibfnamefont {D.~I.}\ \bibnamefont {Schuster}}, \bibinfo {author}
  {\bibfnamefont {J.}~\bibnamefont {Majer}}, \bibinfo {author} {\bibfnamefont
  {A.}~\bibnamefont {Blais}}, \bibinfo {author} {\bibfnamefont {M.~H.}\
  \bibnamefont {Devoret}}, \bibinfo {author} {\bibfnamefont {S.~M.}\
  \bibnamefont {Girvin}}, \ and\ \bibinfo {author} {\bibfnamefont {R.~J.}\
  \bibnamefont {Schoelkopf}},\ }\href@noop {} {\bibfield  {journal} {\bibinfo
  {journal} {Phys. Rev. A}\ }\textbf {\bibinfo {volume} {76}},\ \bibinfo
  {pages} {042319} (\bibinfo {year} {2007})}\BibitemShut {NoStop}%
\bibitem [{\citenamefont {Mi}\ \emph {et~al.}(2017{\natexlab{b}})\citenamefont
  {Mi}, \citenamefont {Cady}, \citenamefont {Zajac}, \citenamefont {Deelman},\
  and\ \citenamefont {Petta}}]{Mi_Science_2016}%
  \BibitemOpen
  \bibfield  {author} {\bibinfo {author} {\bibfnamefont {X.}~\bibnamefont
  {Mi}}, \bibinfo {author} {\bibfnamefont {J.~V.}\ \bibnamefont {Cady}},
  \bibinfo {author} {\bibfnamefont {D.~M.}\ \bibnamefont {Zajac}}, \bibinfo
  {author} {\bibfnamefont {P.~W.}\ \bibnamefont {Deelman}}, \ and\ \bibinfo
  {author} {\bibfnamefont {J.~R.}\ \bibnamefont {Petta}},\ }\href@noop {}
  {\bibfield  {journal} {\bibinfo  {journal} {Science}\ }\textbf {\bibinfo
  {volume} {355}},\ \bibinfo {pages} {156} (\bibinfo {year}
  {2017}{\natexlab{b}})}\BibitemShut {NoStop}%
\bibitem [{\citenamefont {Burkard}\ and\ \citenamefont
  {Petta}(2016)}]{PhysRevB.94.195305}%
  \BibitemOpen
  \bibfield  {author} {\bibinfo {author} {\bibfnamefont {G.}~\bibnamefont
  {Burkard}}\ and\ \bibinfo {author} {\bibfnamefont {J.~R.}\ \bibnamefont
  {Petta}},\ }\href@noop {} {\bibfield  {journal} {\bibinfo  {journal} {Phys.
  Rev. B}\ }\textbf {\bibinfo {volume} {94}},\ \bibinfo {pages} {195305}
  (\bibinfo {year} {2016})}\BibitemShut {NoStop}%
\bibitem [{\citenamefont {Wallraff}\ \emph {et~al.}(2004)\citenamefont
  {Wallraff}, \citenamefont {Schuster}, \citenamefont {Blais}, \citenamefont
  {Frunzio}, \citenamefont {Huang}, \citenamefont {Majer}, \citenamefont
  {Kumar}, \citenamefont {Girvin},\ and\ \citenamefont
  {Schoelkopf}}]{Strong_Coupling_CooperPair}%
  \BibitemOpen
  \bibfield  {author} {\bibinfo {author} {\bibfnamefont {A.}~\bibnamefont
  {Wallraff}}, \bibinfo {author} {\bibfnamefont {D.~I.}\ \bibnamefont
  {Schuster}}, \bibinfo {author} {\bibfnamefont {A.}~\bibnamefont {Blais}},
  \bibinfo {author} {\bibfnamefont {L.}~\bibnamefont {Frunzio}}, \bibinfo
  {author} {\bibfnamefont {R.-S.}\ \bibnamefont {Huang}}, \bibinfo {author}
  {\bibfnamefont {J.}~\bibnamefont {Majer}}, \bibinfo {author} {\bibfnamefont
  {S.}~\bibnamefont {Kumar}}, \bibinfo {author} {\bibfnamefont {S.~M.}\
  \bibnamefont {Girvin}}, \ and\ \bibinfo {author} {\bibfnamefont {R.~J.}\
  \bibnamefont {Schoelkopf}},\ }\href@noop {} {\bibfield  {journal} {\bibinfo
  {journal} {Nature (London)}\ }\textbf {\bibinfo {volume} {431}},\ \bibinfo
  {pages} {162} (\bibinfo {year} {2004})}\BibitemShut {NoStop}%
\bibitem [{Sup()}]{Supplement}%
  \BibitemOpen
  \href@noop {} {\bibinfo  {journal} {See Supplemental Material at [URL will be
  inserted by publisher] for a large-scale DQD charge stability diagram,
  microwave spectroscopy of the excited valley states, and details of the
  theory fits in Fig.~2 and Fig.~4}\ }\BibitemShut {NoStop}%
\bibitem [{\citenamefont {Landau}(1932)}]{Landau_1932}%
  \BibitemOpen
\bibfield  {journal} {  }\bibfield  {author} {\bibinfo {author} {\bibfnamefont
  {L.}~\bibnamefont {Landau}},\ }\href@noop {} {\bibfield  {journal} {\bibinfo
  {journal} {Phys. Z. Sowjetunion}\ }\textbf {\bibinfo {volume} {2}},\ \bibinfo
  {pages} {46} (\bibinfo {year} {1932})}\BibitemShut {NoStop}%
\bibitem [{\citenamefont {Zener}(1932)}]{Zener_1932}%
  \BibitemOpen
  \bibfield  {author} {\bibinfo {author} {\bibfnamefont {C.}~\bibnamefont
  {Zener}},\ }\href@noop {} {\bibfield  {journal} {\bibinfo  {journal} {Proc.
  R. Soc. London, Ser. A}\ }\textbf {\bibinfo {volume} {137}},\ \bibinfo
  {pages} {696} (\bibinfo {year} {1932})}\BibitemShut {NoStop}%
\bibitem [{\citenamefont {St\"uckelberg}(1932)}]{Stuckelberg_1932}%
  \BibitemOpen
  \bibfield  {author} {\bibinfo {author} {\bibfnamefont {E.}~\bibnamefont
  {St\"uckelberg}},\ }\href@noop {} {\bibfield  {journal} {\bibinfo  {journal}
  {Helv. Phys. Acta}\ }\textbf {\bibinfo {volume} {5}},\ \bibinfo {pages} {36}
  (\bibinfo {year} {1932})}\BibitemShut {NoStop}%
\bibitem [{\citenamefont {Majorana}(1932)}]{Majorana_1932}%
  \BibitemOpen
  \bibfield  {author} {\bibinfo {author} {\bibfnamefont {E.}~\bibnamefont
  {Majorana}},\ }\href@noop {} {\bibfield  {journal} {\bibinfo  {journal}
  {Nuovo Cimento}\ }\textbf {\bibinfo {volume} {9}},\ \bibinfo {pages} {43}
  (\bibinfo {year} {1932})}\BibitemShut {NoStop}%
\bibitem [{\citenamefont {Oliver}\ \emph {et~al.}(2005)\citenamefont {Oliver},
  \citenamefont {Yu}, \citenamefont {Lee}, \citenamefont {Berggren},
  \citenamefont {Levitov},\ and\ \citenamefont
  {Orlando}}]{Oliver_Science_2005}%
  \BibitemOpen
  \bibfield  {author} {\bibinfo {author} {\bibfnamefont {W.~D.}\ \bibnamefont
  {Oliver}}, \bibinfo {author} {\bibfnamefont {Y.}~\bibnamefont {Yu}}, \bibinfo
  {author} {\bibfnamefont {J.~C.}\ \bibnamefont {Lee}}, \bibinfo {author}
  {\bibfnamefont {K.~K.}\ \bibnamefont {Berggren}}, \bibinfo {author}
  {\bibfnamefont {L.~S.}\ \bibnamefont {Levitov}}, \ and\ \bibinfo {author}
  {\bibfnamefont {T.~P.}\ \bibnamefont {Orlando}},\ }\href@noop {} {\bibfield
  {journal} {\bibinfo  {journal} {Science}\ }\textbf {\bibinfo {volume}
  {310}},\ \bibinfo {pages} {1653} (\bibinfo {year} {2005})}\BibitemShut
  {NoStop}%
\bibitem [{\citenamefont {Stehlik}\ \emph {et~al.}(2012)\citenamefont
  {Stehlik}, \citenamefont {Dovzhenko}, \citenamefont {Petta}, \citenamefont
  {Johansson}, \citenamefont {Nori}, \citenamefont {Lu},\ and\ \citenamefont
  {Gossard}}]{Stehlik_PRB_2012}%
  \BibitemOpen
  \bibfield  {author} {\bibinfo {author} {\bibfnamefont {J.}~\bibnamefont
  {Stehlik}}, \bibinfo {author} {\bibfnamefont {Y.}~\bibnamefont {Dovzhenko}},
  \bibinfo {author} {\bibfnamefont {J.~R.}\ \bibnamefont {Petta}}, \bibinfo
  {author} {\bibfnamefont {J.~R.}\ \bibnamefont {Johansson}}, \bibinfo {author}
  {\bibfnamefont {F.}~\bibnamefont {Nori}}, \bibinfo {author} {\bibfnamefont
  {H.}~\bibnamefont {Lu}}, \ and\ \bibinfo {author} {\bibfnamefont {A.~C.}\
  \bibnamefont {Gossard}},\ }\href@noop {} {\bibfield  {journal} {\bibinfo
  {journal} {Phys. Rev. B}\ }\textbf {\bibinfo {volume} {86}},\ \bibinfo
  {pages} {121303} (\bibinfo {year} {2012})}\BibitemShut {NoStop}%
\bibitem [{\citenamefont {Forster}\ \emph {et~al.}(2014)\citenamefont
  {Forster}, \citenamefont {Petersen}, \citenamefont {Manus}, \citenamefont
  {H\"anggi}, \citenamefont {Schuh}, \citenamefont {Wegscheider}, \citenamefont
  {Kohler},\ and\ \citenamefont {Ludwig}}]{Forster_PRL_2017}%
  \BibitemOpen
  \bibfield  {author} {\bibinfo {author} {\bibfnamefont {F.}~\bibnamefont
  {Forster}}, \bibinfo {author} {\bibfnamefont {G.}~\bibnamefont {Petersen}},
  \bibinfo {author} {\bibfnamefont {S.}~\bibnamefont {Manus}}, \bibinfo
  {author} {\bibfnamefont {P.}~\bibnamefont {H\"anggi}}, \bibinfo {author}
  {\bibfnamefont {D.}~\bibnamefont {Schuh}}, \bibinfo {author} {\bibfnamefont
  {W.}~\bibnamefont {Wegscheider}}, \bibinfo {author} {\bibfnamefont
  {S.}~\bibnamefont {Kohler}}, \ and\ \bibinfo {author} {\bibfnamefont
  {S.}~\bibnamefont {Ludwig}},\ }\href@noop {} {\bibfield  {journal} {\bibinfo
  {journal} {Phys. Rev. Lett.}\ }\textbf {\bibinfo {volume} {112}},\ \bibinfo
  {pages} {116803} (\bibinfo {year} {2014})}\BibitemShut {NoStop}%
\bibitem [{\citenamefont {Shevchenko}\ \emph {et~al.}(2010)\citenamefont
  {Shevchenko}, \citenamefont {Ashhab},\ and\ \citenamefont
  {Nori}}]{Shevchenko_Review_2010}%
  \BibitemOpen
  \bibfield  {author} {\bibinfo {author} {\bibfnamefont {S.}~\bibnamefont
  {Shevchenko}}, \bibinfo {author} {\bibfnamefont {S.}~\bibnamefont {Ashhab}},
  \ and\ \bibinfo {author} {\bibfnamefont {F.}~\bibnamefont {Nori}},\
  }\href@noop {} {\bibfield  {journal} {\bibinfo  {journal} {Phys. Rep.}\
  }\textbf {\bibinfo {volume} {492}},\ \bibinfo {pages} {1 } (\bibinfo {year}
  {2010})}\BibitemShut {NoStop}%
\bibitem [{\citenamefont {Bienfait}\ \emph {et~al.}(2016)\citenamefont
  {Bienfait}, \citenamefont {Pla}, \citenamefont {Kubo}, \citenamefont {Zhou},
  \citenamefont {Stern}, \citenamefont {Lo}, \citenamefont {Weis},
  \citenamefont {Schenkel}, \citenamefont {Vion}, \citenamefont {Esteve},
  \citenamefont {Morton},\ and\ \citenamefont {Bertet}}]{Bienfait_Nature_2016}%
  \BibitemOpen
  \bibfield  {author} {\bibinfo {author} {\bibfnamefont {A.}~\bibnamefont
  {Bienfait}}, \bibinfo {author} {\bibfnamefont {J.~J.}\ \bibnamefont {Pla}},
  \bibinfo {author} {\bibfnamefont {Y.}~\bibnamefont {Kubo}}, \bibinfo {author}
  {\bibfnamefont {X.}~\bibnamefont {Zhou}}, \bibinfo {author} {\bibfnamefont
  {M.}~\bibnamefont {Stern}}, \bibinfo {author} {\bibfnamefont {C.~C.}\
  \bibnamefont {Lo}}, \bibinfo {author} {\bibfnamefont {C.~D.}\ \bibnamefont
  {Weis}}, \bibinfo {author} {\bibfnamefont {T.}~\bibnamefont {Schenkel}},
  \bibinfo {author} {\bibfnamefont {D.}~\bibnamefont {Vion}}, \bibinfo {author}
  {\bibfnamefont {D.}~\bibnamefont {Esteve}}, \bibinfo {author} {\bibfnamefont
  {J.~J.~L.}\ \bibnamefont {Morton}}, \ and\ \bibinfo {author} {\bibfnamefont
  {P.}~\bibnamefont {Bertet}},\ }\href@noop {} {\bibfield  {journal} {\bibinfo
  {journal} {Nature (London)}\ }\textbf {\bibinfo {volume} {531}},\ \bibinfo
  {pages} {74} (\bibinfo {year} {2016})}\BibitemShut {NoStop}%
\bibitem [{\citenamefont {Stehlik}\ \emph {et~al.}(2016)\citenamefont
  {Stehlik}, \citenamefont {Liu}, \citenamefont {Eichler}, \citenamefont
  {Hartke}, \citenamefont {Mi}, \citenamefont {Gullans}, \citenamefont
  {Taylor},\ and\ \citenamefont {Petta}}]{Stehlik_PRX}%
  \BibitemOpen
  \bibfield  {author} {\bibinfo {author} {\bibfnamefont {J.}~\bibnamefont
  {Stehlik}}, \bibinfo {author} {\bibfnamefont {Y.~Y.}\ \bibnamefont {Liu}},
  \bibinfo {author} {\bibfnamefont {C.}~\bibnamefont {Eichler}}, \bibinfo
  {author} {\bibfnamefont {T.~R.}\ \bibnamefont {Hartke}}, \bibinfo {author}
  {\bibfnamefont {X.}~\bibnamefont {Mi}}, \bibinfo {author} {\bibfnamefont
  {M.~J.}\ \bibnamefont {Gullans}}, \bibinfo {author} {\bibfnamefont {J.~M.}\
  \bibnamefont {Taylor}}, \ and\ \bibinfo {author} {\bibfnamefont {J.~R.}\
  \bibnamefont {Petta}},\ }\href@noop {} {\bibfield  {journal} {\bibinfo
  {journal} {Phys. Rev. X}\ }\textbf {\bibinfo {volume} {6}},\ \bibinfo {pages}
  {041027} (\bibinfo {year} {2016})}\BibitemShut {NoStop}%
\bibitem [{\citenamefont {Koski}\ \emph {et~al.}()\citenamefont {Koski},
  \citenamefont {Landig}, \citenamefont {Palyi}, \citenamefont {Scarlino},
  \citenamefont {Reichl}, \citenamefont {Wegscheider}, \citenamefont {Burkard},
  \citenamefont {Wallraff}, \citenamefont {Ensslin},\ and\ \citenamefont
  {Ihn}}]{Koski_Arxiv_2018}%
  \BibitemOpen
  \bibfield  {author} {\bibinfo {author} {\bibfnamefont {J.~V.}\ \bibnamefont
  {Koski}}, \bibinfo {author} {\bibfnamefont {A.~J.}\ \bibnamefont {Landig}},
  \bibinfo {author} {\bibfnamefont {A.}~\bibnamefont {Palyi}}, \bibinfo
  {author} {\bibfnamefont {P.}~\bibnamefont {Scarlino}}, \bibinfo {author}
  {\bibfnamefont {C.}~\bibnamefont {Reichl}}, \bibinfo {author} {\bibfnamefont
  {W.}~\bibnamefont {Wegscheider}}, \bibinfo {author} {\bibfnamefont
  {G.}~\bibnamefont {Burkard}}, \bibinfo {author} {\bibfnamefont
  {A.}~\bibnamefont {Wallraff}}, \bibinfo {author} {\bibfnamefont
  {K.}~\bibnamefont {Ensslin}}, \ and\ \bibinfo {author} {\bibfnamefont
  {T.}~\bibnamefont {Ihn}},\ }\href@noop {} {\bibinfo  {journal}
  {arXiv:1802.03810}\ }\BibitemShut {NoStop}%
\bibitem [{\citenamefont {Kohler}(2017)}]{Kohler_PRL_2017}%
  \BibitemOpen
\bibfield  {journal} {  }\bibfield  {author} {\bibinfo {author} {\bibfnamefont
  {S.}~\bibnamefont {Kohler}},\ }\href@noop {} {\bibfield  {journal} {\bibinfo
  {journal} {Phys. Rev. Lett.}\ }\textbf {\bibinfo {volume} {119}},\ \bibinfo
  {pages} {196802} (\bibinfo {year} {2017})}\BibitemShut {NoStop}%
\bibitem [{\citenamefont {Basset}\ \emph {et~al.}(2014)\citenamefont {Basset},
  \citenamefont {Stockklauser}, \citenamefont {Jarausch}, \citenamefont {Frey},
  \citenamefont {Reichl}, \citenamefont {Wegscheider}, \citenamefont
  {Wallraff}, \citenamefont {Ensslin},\ and\ \citenamefont
  {Ihn}}]{Walraff_ChargeNoise_APL}%
  \BibitemOpen
  \bibfield  {author} {\bibinfo {author} {\bibfnamefont {J.}~\bibnamefont
  {Basset}}, \bibinfo {author} {\bibfnamefont {A.}~\bibnamefont
  {Stockklauser}}, \bibinfo {author} {\bibfnamefont {D.-D.}\ \bibnamefont
  {Jarausch}}, \bibinfo {author} {\bibfnamefont {T.}~\bibnamefont {Frey}},
  \bibinfo {author} {\bibfnamefont {C.}~\bibnamefont {Reichl}}, \bibinfo
  {author} {\bibfnamefont {W.}~\bibnamefont {Wegscheider}}, \bibinfo {author}
  {\bibfnamefont {A.}~\bibnamefont {Wallraff}}, \bibinfo {author}
  {\bibfnamefont {K.}~\bibnamefont {Ensslin}}, \ and\ \bibinfo {author}
  {\bibfnamefont {T.}~\bibnamefont {Ihn}},\ }\href@noop {} {\bibfield
  {journal} {\bibinfo  {journal} {Appl. Phys. Lett.}\ }\textbf {\bibinfo
  {volume} {105}},\ \bibinfo {eid} {063105} (\bibinfo {year}
  {2014})}\BibitemShut {NoStop}%
\end{thebibliography}%


\end{document}